\documentclass[preprint,showpacs,amsmath]{revtex4}
\usepackage[dvips]{graphicx}
\usepackage{amsmath}
\usepackage{xspace}
\usepackage{color}

\begin{document}

\title{
Deviation from power law of the global
seismic moment distribution
}

\author{Isabel Serra$^1$, \'Alvaro Corral$^{1,2}$}

\affiliation{
$^1$Centre de Recerca Matem\`atica, Campus de Bellaterra, Edifici C, E-08193 Barcelona, Spain\\
$^2$Departament de Matem\`atiques, Universitat Aut\`onoma de Barcelona, E-08193 Barcelona, Spain\\
}
\date{\today}

\begin{abstract}
The distribution of seismic moment is of capital interest to evaluate earthquake hazard,
in particular regarding the most extreme events.
We make use of likelihood-ratio tests to compare the
simple Gutenberg-Richter power-law distribution with
two statistical models that incorporate an exponential tail:
the so-called tapered Gutenberg-Richter and the truncated gamma,
when fitted to the global CMT earthquake catalog.
The outcome is that the truncated gamma model outperforms the
other two models.
If simulated samples of the truncated gamma are reshuffled
in order to mimic the time occurrence of the order statistics
of the empirical data,
this model turns out to be able
to explain the empirical data both before and
after the great Sumatra-Andaman earthquake of 2004.
%
%
%
\end{abstract}


\maketitle




The Gutenberg-Richter (GR) law is not only of fundamental importance in statistical seismology
\cite{Utsu_GR}
but also a cornerstone of non-linear geophysics \cite{Malamud_hazards}
and complex-systems science \cite{Bak_book}.
It simply states that, for a given region,
the magnitudes of earthquakes follow an exponential probability distribution.
As the (scalar) seismic moment is an exponential function of magnitude, when
the GR law is expressed in terms of the former variable,
it translates into a power-law distribution \cite{Knopoff_Kagan77,Corral_Lacidogna},
i.e.,
\begin{equation}
f(M) \propto \frac 1 {M^{1+\beta}},
\label{laprimerapowerlaw}
\end{equation}
with $M$ seismic moment, $f(M)$ its probability density,
(fulfilling $\int_{\forall M} f(M) dM =1$),
the sign ``$\propto$'' denoting proportionality,
and the exponent $1+\beta$ taking values close to 1.65.
This simple description provides rather good fits of available
data in many cases \cite{Kagan_pageoph99,Kagan_gji02,Corral_Deluca,Kagan_book},
with, remarkably, only one free parameter, $\beta$.
A totally equivalent characterization of the distribution uses
the survivor function (or complementary cumulative distribution),
defined as $S(M)=\int_M^\infty f(M') dM'$, for which the GR power law
takes the form $S(M) \propto 1/M^\beta$.

The power-law distribution has important physical implications,
as it suggests an origin from a critical branching process
or a self-organized-critical system \cite{Bak_book,Vere_Jones76,Main}.
Nevertheless, it presents also
some conceptual difficulties,
due to the fact that the mean value $\langle M \rangle$
provided by the distribution turns out to be infinite \cite{Knopoff_Kagan77}.
These elementary considerations imply that the
GR law cannot be naively extended to arbitrarily large values of $M$,
and one needs to introduce additional parameters to
describe the tail of the distribution, coming presumably from
finite-size effects.
However, a big problem is that
the change from power law to a faster decay seems to take
place at the highest values of $M$ that have been observed,
for which the statistics are very poor \cite{Zoller_grl}.

Kagan \cite{Kagan_gji02} has enumerated the requirements that an
extension of the GR law should fulfill; in particular, he
considered, among other:
(i) the so called tapered (Tap) Gutenberg-Richter distribution
(also called Kagan distribution \cite{Vere_Jones_gji}),
with a survivor function given by
$S_{tap}(M) \propto e^{-M/\theta}/M ^\beta$ and
(ii) the (left-) truncated gamma (TrG) distribution,
for which the density is
$f_{trg}(M) \propto e^{-M/\theta}/M ^{1+\beta}$.
Note that both expressions have essentially the same functional form,
but the former refers to the survivor function and the later
to the density.
As, in general, $f(M) = -dS(M) / d M$,
differentiation of $S_{tap}(M)$ in (i) shows the difference between both
distributions.
In any case, parameter $\theta$ represents a crossover value of seismic moment, signaling a
transition from power law to exponential decay; so, $\theta$ gives the scale
of the finite-size effects on the seismic moment. The corresponding
value of (moment) magnitude (sometimes called corner magnitude)
can be obtained from $m_c=\frac 2 3
\left( \log_{10} \theta - 9.1\right)$, when the seismic moment is
measured in N$\cdot$m \cite{Kanamori_77,Kanamori_rpp}.

Kagan \cite{Kagan_gji02} also argues that available
seismic catalogs do not allow the reliable estimation of $\theta$,
except in the global case (or for large subsets of this case),
in particular, he recommends the use of the
centroid moment tensor (CMT) catalog \cite{Ekstrom2012}.
From
his analysis of global seismicity, and comparing the values
of the likelihoods, Kagan \cite{Kagan_gji02} concludes that the
tapered GR distribution gives a slightly better fit than the truncated gamma
distribution, for which in addition the estimation procedure is more
involving. In any case, the $\beta-$value seems to be universal (at
variance with $\theta$), see also Refs. \cite{Kagan_book,Godano_pingue,Kagan_tectono10}.

Nevertheless, the data analyzed by Kagan \cite{Kagan_gji02}, from 1977 to 1999,
comprises a period of relatively low global seismic activity, with no
event above magnitude 8.5;
in contrast, the period 1950 -- 1965 witnessed 7 of such events
\cite{Lay}.
Starting with the great Sumatra-Andaman earthquake of 2004,
and following since then with 4 more earthquakes with $m \ge 8.5$,
the current period seems to correspond to the past higher levels of activity
(up to the time of submitting this letter).
%
%
Main {\it et al.} and Bell {\it et al.}
\cite{Main_ng,Bell_Naylor_Main} have re-examined the problem of the
seismic moment distribution including recent global data (shallow
events only). Using a Bayesian information criterion (BIC),
Bell {\it et al.} \cite{Bell_Naylor_Main}
compare the plain GR power law with the tapered GR distribution, and
conclude that, although the tapered GR gives a significantly better
fit before the 2004 Sumatra event, the occurrence of this changes
the balance of the BIC statistics, making the GR power law more
suitable; that is, the power law is more parsimonious, or simply, is
enough for describing global shallow seismicity when the recent
mega-earthquakes are included in the data.
Similar results have been published in Ref. \cite{Geist_Parsons_NH}.

In this paper we revisit the problem with more recent data,
using other statistical tools, reaching
{somewhat} different conclusions.
As in the mentioned papers, we analyze the
global CMT catalog \cite{Ekstrom2012}, in our case for the period
between 1 January 1977 and 31 October
2013, with
the values of the seismic moment converted into N$\cdot$m
($1$ dyn$\cdot$cm
$=10^{-7}$ N$\cdot$m).
We restrict to shallow events (depth $ < 70 $ km)
and in order to avoid incompleteness, to magnitude $m > 5.75$
(equivalent to $M>
5.3 \cdot 10^{17}$ N$\cdot$m),
as in Refs. \cite{Main_ng,Bell_Naylor_Main}.
This yields $6150$ events.

As statistical tools, we use maximum likelihood estimation (MLE) for
fitting, and likelihood ratio (LR) tests for comparison of different
fits. Model selection tests based on the likelihood ratio have the
advantage that the ratio is invariant with respect to changes of
variables (if these are one-to-one \cite{Pawitan2001}). Moreover,
for comparing the fit of models in pairs, LR {test} is preferable in front
of the computation of differences in BIC or AIC (Akaike information
criterion), as the test relies on the fact that the distribution of
the LR is known, under a suitable null hypothesis
{(although the log-likelihood-ratio is equal to the difference of BIC
or AIC when the number of parameters of the two models is the same).}




Maximum likelihood estimation is the best-accepted method in order
to fit probability distributions, as it yields estimators which are
invariant under re-parameterizations, and
which are asymptotically unbiased and efficient for regular models,
in particular for exponential families
\cite{Pawitan2001}.
When maximum likelihood is used under a wrong model,
what one finds is the closest model to the true distribution
in terms of the Kullback-Leibler distance \cite{Pawitan2001}.
%
%
In order to perform MLE it is necessary
to specify the densities of the distributions, including the normalization factors.
In our case,
all distributions are defined for $M$ above the completeness threshold $a$,
i.e., for $M > a$, being zero otherwise
(as mentioned above, $a$ is fixed to $5.3 \cdot 10^{17}$ N$\cdot$m).
For the power-law (PL) distribution
(which yields the GR law for the distribution of $M$)
Eq. (\ref{laprimerapowerlaw}) reads
$$
f_{pl}(M; \beta)
=\frac {\beta} a \left( \frac a M\right)^{1+\beta},
$$
with $\beta > 0$.
For the tapered Gutenberg-Richter, 
$$
f_{tap}(M; \beta, \theta)=\left[\frac \beta a \left(\frac a M\right)^{1+\beta} + \frac 1 \theta \left(\frac a M\right)^{\beta}\right] e^{-(M-a)/\theta},
$$
with $\beta >0$ and $\theta >0$.
And for the left-truncated (and extended to $\beta >0$) gamma 
distribution;
$$
f_{trg}(M; \beta,\theta) = \frac 1 {\theta \Gamma(-\beta,a/\theta)} \left(\frac \theta M \right)^{1+\beta} e^{-M/\theta},
$$
with $-\infty <\beta < \infty$ and $\theta >0$,
and with $\Gamma(\gamma,z)= \int_z^\infty x^{\gamma-1} e^{-x}dx$
the upper incomplete gamma function, defined for $z>0$ when $\gamma <0$.
%
%
We summarize the parameterization of the densities as $f(M;\Theta)$,
where $\Theta=\{\beta,\theta\}$ for the Tap and TrG distributions
and $\Theta=\beta$ for the power law.
Note that for the 
TrG distribution, it is clear that
the exponent $\beta$ is a shape parameter and
$\theta$ is a scale parameter;
in fact, these parameters play the same role in the Tap distribution,
which turns out to be a mixture of two truncated gamma distributions,
one with shape parameter $\beta$ and the other with $\beta-1$,
but with common scale parameter $\theta$.
In contrast, the power law 
lacks a scale parameter.
{In all cases the completeness threshold $a$ is a truncation parameter, but it
is kept fixed and is not a free parameter, therefore.}


The knowledge of the probability densities allows the direct
computation of the likelihood function as $ L(\Theta)= \prod_{i=1}^N
f(M_i;\Theta), $ where $M_i$ are the $N$ observational values of the
seismic moment. Maximization of the likelihood function with respect
the values of the parameters leads to the maximum-likelihood
estimation $\hat \Theta$ of these parameters, with $\hat L = L(\hat
\Theta)$ the value of the likelihood at its maximum.
We perform the MLE for the 3
models, obtaining, for the complete datasets, the values reported in
Table \ref{taula}.

%

\begin{table}[h]
\begin{centering}
\begin{tabular}{|c|ccccc|}
\hline
& & $\hat \beta$ & $\hat \theta$ (N$\cdot$m) & $\hat m_c$ & $l$ ($M$ in N$\cdot$m)\tabularnewline
\cline{3-6}
PL & MLE & 0.685 & $\infty$ & $\infty$ & -268466.609 
\tabularnewline
& s.e. & 0.009 & & & \tabularnewline
\hline
Tap & MLE & 0.684 & 3.3\:$10^{22}$ &8.94 & -268465.315 
\tabularnewline
& s.e. & 0.009 & 2.6\:$10^{22}$ & 0.23 & \tabularnewline
\hline
TrG & MLE & 0.681 & 6.7\:$10^{22}$ &9.15 & -268464.844 
\tabularnewline
& s.e. & 0.009 & 6.6\:$10^{22}$ & 0.27 & \tabularnewline
\hline
\end{tabular}
\par\end{centering}
\centering{}\caption{
Maximum likelihood estimation of the parameters
with their standard errors (s.e.)
and maximum value of the log-likelihood function, $l=\ln \hat L$
when the PL, Tap, and TrG distributions
{are fitted to the seismic moment of shallow CMT earthquakes},
using the whole data set ($N=6150$).
The standard error for $\hat \beta$ and $\hat \theta$
is computed from the Fisher information matrix and
corresponds to one standard deviation of the distribution of {each} parameter.
The standard error for $\hat m_c$ is computed from that of $\hat \theta$
using the delta method \cite{Casella}.
\label{taula}
}
\end{table}

A powerful method for comparison of pairs of models is the likelihood-ratio test,
specially suitable when one model is nested within the other,
which means that the first model is obtained as a special case
of the second one.
This is the case of the power-law distribution
with respect to the other two distributions;
indeed, the power law
is nested both within the Tap and within the truncated gamma, as
taking $\theta \rightarrow \infty$ in any of the two leads to the power-law
distribution. This is easily seen taking into account that
$S_{tap}(M) = (a/M)^\beta e^{-(M-a)/\theta}$,
or just performing the limit in the expression for $f_{tap}(M)$ above.
For the truncated gamma distribution,
when doing the $\theta \rightarrow \infty$ limit in $f_{trg}(M)$
one needs to use that, for $\gamma<0$,
${z^{\gamma}/\Gamma(\gamma,z)\rightarrow -\gamma}$
when $z\rightarrow 0$,
see Ref. \cite{Gamma_expansion} for $\gamma \ne -1, -2, \dots$

Given two probability distributions, $1$ and $2$,
with $1$ nested within $2$,
the likelihood ratio test evaluates $\hat L_{2}/\hat L_{1}$,
where $\hat L_{2}$ is the likelihood (at maximum)
of the ``bigger'' or ``full'' model (either Tap or TrG)
and $\hat L_{1}$ corresponds to the nested or null model
(power law in our case). Taking logarithms we get
the log-likelihood-ratio
%
\[
\mathcal{R}=\ln \frac{\hat L_2}{\hat L_1}= l_{2}-l_{1},
\]
with $ l_{j}=\ln \hat L_j=\sum_{i=1}^{N} \ln f_{j}(M_i;\hat{\Theta}_j) $,
where $f_{i}$ denotes the probability density function of the distribution $j$
for every $j=1,2$, and the MLE corresponds to $\hat \Theta_1=\hat
\beta_1$ and $\hat\Theta_2=\{\hat\beta_2,\hat\theta_2\}$. In order
to compare the fit provided by the two distributions, it is
necessary to characterize the distribution of $\mathcal{R}$.

Let $n_1$ and $n_2$ be the number of free parameters in the models
$1$ and $2$, respectively.
In general, if the models are nested,
and under the null
hypothesis that the data comes from the simpler model, the
probability distribution of the statistic $2 \mathcal{R}$ in the limit
$N\rightarrow \infty$ is
a chi-squared distribution
with degrees of freedom equal to $n_2-n_1 > 0$.
So,
\begin{equation}
2 \mathcal{R}>3.84
\end{equation}
with a level of risk equal to 0.05.
{Note that the chi-squared distribution
provides a penalty for
``model complexity''
as the {``width''} of the distribution is given directly by
the number of the degrees of freedom.}
This likelihood ratio test
constitutes the best option to choose among models 1 and 2, in the
sense that it has a convergence to its asymptotic distribution
faster
than any other test \cite{mccullagh1986}.
The null and
alternative hypotheses correspond to accept model $1$ or $2$,
respectively, although the acceptance of model $1$ does not imply the rejection
of $2$, it is simply that
the ``full'' model $2$
does not bring any significant improvement with respect the simpler
model $1$.

{However, when the nesting of distribution $1$ within $2$ takes place
in such a way that the space of parameters of the former one
lies within a boundary of the space of parameters of distribution $2$,
the approach just explained for
the asymptotic distribution of $2 \mathcal{R}$
is not valid \cite{Self1987,Geyer1994}.
This is the case when testing both the Tap or the TrG distributions
in front of the power-law distribution,
as the $\theta \rightarrow \infty$ limit of the latter
corresponds to the boundary of the parameter space of the two other distributions.
But the asymptotic theory of Refs. \cite{Self1987,Geyer1994}
is also invalid, as the power-law distribution lacks
the necessary regularity conditions,
due to the divergence of their moments.
This illustrates part of the difficulties
of performing proper model selection when
fractal-like distributions are involved
\cite{Kagan_calcutta}. 
In order to obtain the distribution of $2\mathcal{R}$ and from there
the $p-$values of the LR tests,
we are left to the simulation of the null hypothesis.
We advance that the results seem to indicate that the distribution of $2\mathcal{R}$,
for high percentiles, is chi-square with one degree of freedom,
but we lack a theoretical support for this fact.}

\begin{figure}[t!]
\begin{center}
\centerline{\includegraphics[width=9cm]{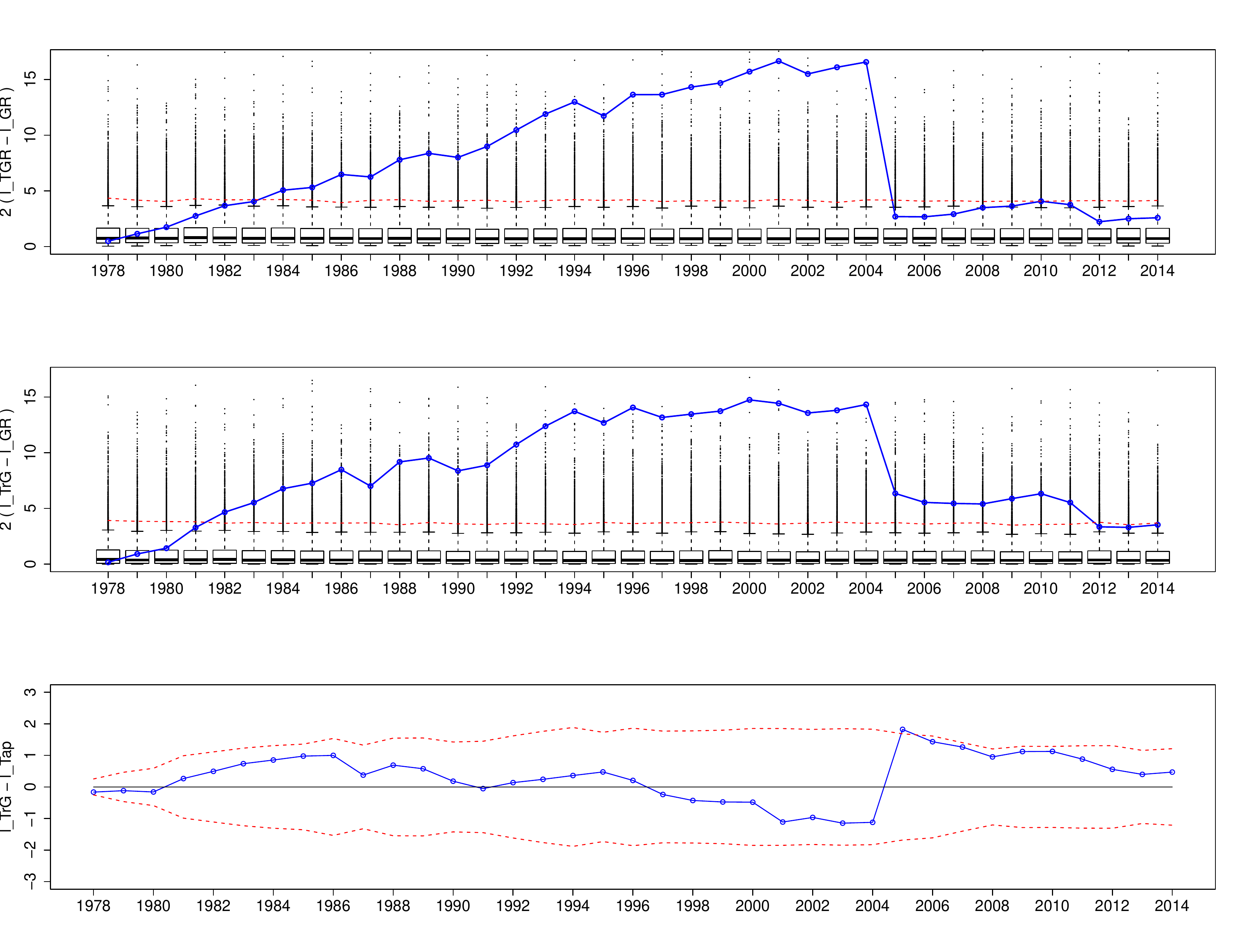}}
\caption{ Results of likelihood ratio tests for nested and
non-nested models. The points denote the value of the statistic $2
\mathcal{R}$ or $\mathcal{R}$ (depending on the test) and the dashed
lines show the critical value of the corresponding test {(at level
0.05)}. {For the nested case, boxplots show the distribution of
$2\mathcal{R}$ for 10000 simulations of the power-law null
hypothesis, from which the critical value is computed.} The abscissa
corresponds to the ending point of a time window starting always in
1 Jan 1977. Note that the year is considered a continuous variable
(not a categorical variable), so, the time window ending on 31 Dec
2004 takes value $2004.99\dots \simeq 2005$. (a) Tap distribution
versus power law. (b) Truncated gamma versus power law. (c)
Truncated gamma versus Tap {(non-nested case)}.
\label{graf:densities}}
\end{center}
\end{figure}

Let us proceed, using this {method},
by comparing the performance of the power-law and Tap fits
when applied to the global shallow seismic activity,
for time windows starting always in 1977 and ending in the successive times
indexed by the abscissa in Fig. \ref{graf:densities}(a)
(as in Ref. \cite{Bell_Naylor_Main}).
The log-likelihood-ratio of these fits (times 2),
is shown in the figure
together with the critical region of the test. In agreement with
Bell {\it et al.} \cite{Bell_Naylor_Main}, we find that: (i) the
power-law fit can be safely rejected in front of the Tap
distribution for any time window ending between 1980 and before
2004; and (ii) the results change drastically after the occurrence
of the great 2004 Sumatra earthquake,
{for which the power law cannot be rejected at the 0.05 level.
So, for parsimony reasons, the power law becomes preferable in front of the
Tap distribution for time windows
ending later than 2004.
The fact that the Tap distribution cannot be distinguished from the power law
is also in agreement with previous results showing that the contour lines
in the likelihood maps of the Tap distribution are highly non-symmetric
and may be unbounded for smaller levels of risk \cite{Kagan_Schoenberg,Kagan_gji02,Geist_Parsons_NH}.
}


When we compare the power-law fit with the truncated gamma, using
the same test, for the same data, the results are more
significant, see Fig. \ref{graf:densities}(b). The situation
previous to 2004 is the same, with an extremely poor performance of
the power law; but after 2004, despite a big jump again in the value
of the likelihood ratio, the power law continues as being
non-acceptable, at the 0.05 level.
It is only after the great Tohoku earthquake of 2011
that the $p-$value of the test {enters into} the acceptance region,
but keeping values not far from the 0.05 limit.
From here we
conclude that, in order to find an alternative to the power-law
distribution, the truncated gamma distribution is a better
option than the Tap distribution, as it is more clearly distinguishable
from the power law (for this particular data).
Nevertheless, a comparison between
these two distributions (Tap and TrG) seems pertinent.



When the models are not nested, as it happens if we want a direct
comparison between the Tap and the TrG distributions,
%
%
the procedure we use 
is the likelihood ratio test of Vuong for non-nested models \cite{Vuong,Clauset}.
In this case the critical values
depend on the sample size, $N$,
turning out to be that, when $N$ is large, $\mathcal{R}$ is normally
distributed with standard deviation $s \sqrt N$,
where $s$ denotes the standard deviation of the set
$$
\{\ln f_{trg}(M_i;\hat{\beta}_{trg},\hat{\theta}_{trg})-\ln
f_{tap}(M_i;\hat{\beta}_{tap},\hat{\theta}_{tap})\}
$$
for $i=1,\dots, N$ and $\mathcal{R}= l_{trg}-l_{tap}$.
Then, we accept that there exists a
significant difference between the models if
\begin{equation}\label{reg}
|\mathcal{R}|>\:1.96 s \sqrt{N}
\end{equation}
at a level of risk equal to 0.05,
with the model with larger log-likelihood being the preferred one.
The critical value of the test arises because
the null hypothesis is that the mean value of $\mathcal{R}$ is zero (i.e.,
both models are equally close to the true distribution).
Note that the alternative hypothesis corresponds to accept that the
difference between the fit provided by the models is significant. As
the number of parameters is the same for the Tap and TrG models,
their log-likelihood-ratio coincides with the difference in BIC or
AIC, {but,} as mentioned above, the LR test incorporates
a statistical test which specifies the distribution
of the statistic under consideration.
%

Figure \ref{graf:densities}(c) shows the evolution of the
log-likelihood-ratio between the two models, for different time windows (starting always in
1977), together with the critical region of the test given by the
Eq. (\ref{reg}). One can see how the fits provided by the Tap and TrG
distributions do not exhibit significant difference, although the
TrG provides, in general, slightly higher likelihoods. After the
mega-event in 2004 the performance of the TrG fit improves,
approaching the limit of significance.
{This reinforces our conclusion that the TrG distribution is preferred
in front of the Tap and power-law distributions.}

In order to gain further insight, we simulate
random samples following the truncated gamma distribution, with the
parameters $\hat \beta_{trg}$ and $\hat \theta_{trg}$ obtained from
ML estimation of the complete dataset (Table \ref{taula}), with the
same truncation parameter $a$ and number of points ($N=6150$) also.
To avoid that the conclusions depend on the time correlation of magnitudes,
we reshuffle the simulated data in such a way that the occurrence of
the order statistics is the same as for the empirical data; in other
words, the largest simulated event is assigned to take place at the
time of the 2011 Tohoku earthquake (the largest of the CMT catalog
\cite{Bell_Naylor_Main}), the second largest at the time of the 2004
Sumatra event, and so on. In this way, we model earthquake seismic
moments as
arising from a gamma distribution with fixed parameters and with
occurrence times given by the empirical times and with the same
seismic moment correlations
as the empirical data, approximately.

\begin{figure}[t!]
\begin{center}
\centerline{\includegraphics[trim = 0mm 130mm 0mm
10mm,clip=true,width=9cm]{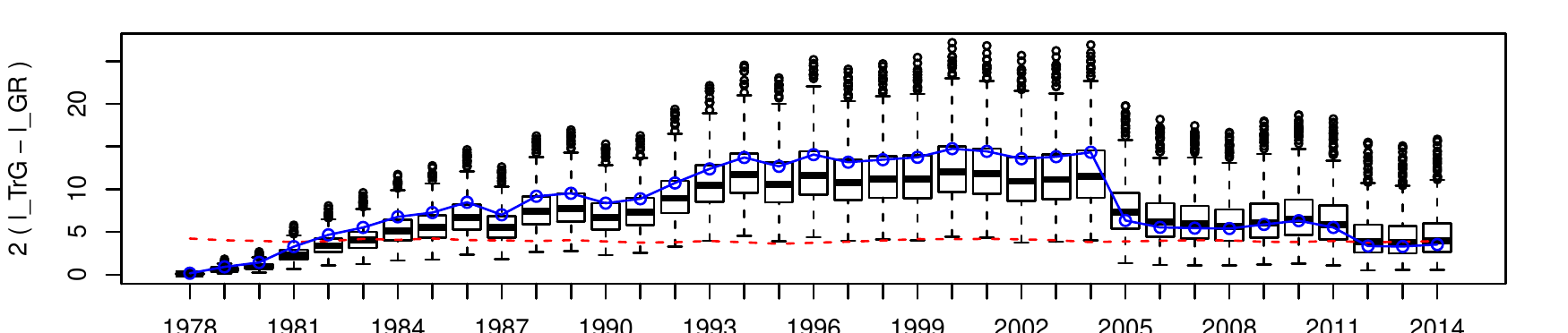}}
\caption{Comparison of the empirical log-likelihood-ratios between
the TrG and power law with those of 1000 simulations of the TrG
distribution, using the final parameters of Table \ref{taula}
{(i.e., $\beta=0.681$ and $m_c=9.15$)}. Simulated seismic moments
are reshuffled as explained in the text {to make the comparison
possible}. Simulation results are displayed using boxplots,
representing the three quartiles of the distribution of
$2\mathcal{R}$.
\label{l_boxplots}}
\end{center}
\end{figure}

\begin{figure}
\begin{center}
\centerline{\includegraphics[trim = 0mm 0mm 0mm
60mm,clip,width=9cm]{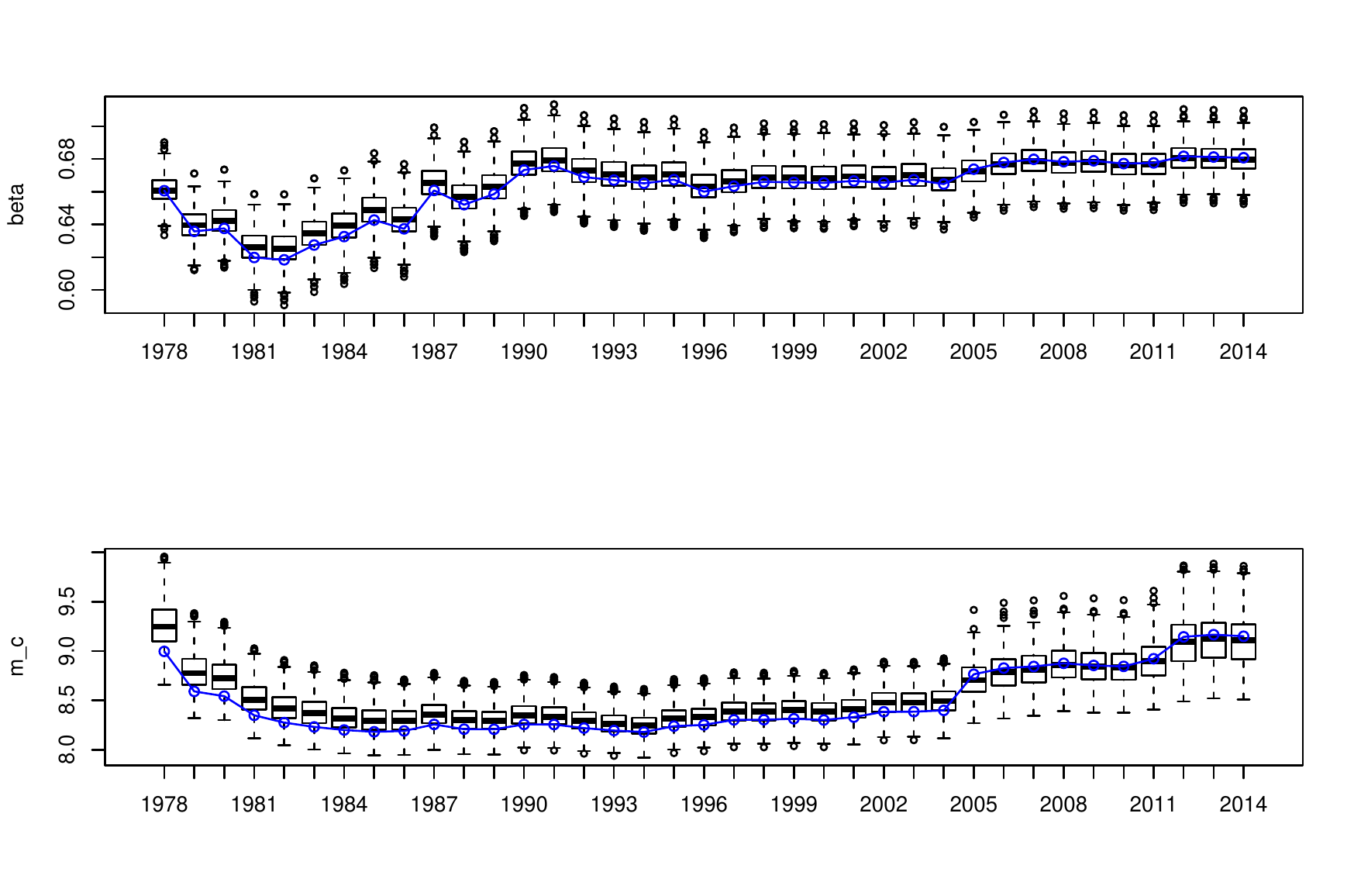}}
\caption{Comparison of the values of the estimated parameters of the
TrG distribution, $\hat \beta_{trg}$ and $\hat m_{c\,trg}$, for the
empirical data and for 1000 simulations of the TrG distribution,
using the final parameters of Table \ref{taula}. Simulated seismic
moments are reshuffled as explained in the text. The different
stability of both parameters is apparent. \label{gr4}}
\end{center}
\end{figure}


We simulate 1000 datasets with $N=6150$ each. The results, displayed
in Fig. \ref{l_boxplots}, show that the behavior of the empirical
data is not atypical in comparison with this gamma modeling. In
nearly all time windows the empirical data lies in between the first
and third quartile of the simulated data, although before 2004 the
empirical values are close to the third quartile whereas after 2004
they lay just below the median. This leads us to compute the
statistics of the jump in the log-likelihood-ratio between 2004 and
2005. The estimated probability of having a jump larger than the
empirical value is around 4.5 \%, which is not far from what one
could {accept} from the gamma modeling explained above.
{Thus, a TrG distribution, with fixed parameters, is able
to reproduce the empirical findings, if the peculiar time ordering of magnitude
of the real events is taken into account.
Notice also that, although the simulated data come from a TrG distribution,
they are not distinguishable from a power law for
about half of the simulations of the last time windows,
as the critical region is close to the median
indicated by the boxplots.}

We can also compare the evolution of the estimated parameters
for the empiral dataset and for the reshuffled TrG simulations, with a good agreement
again, see Fig. \ref{gr4}. There, it is clear that although the exponent $\beta$
reaches very stable values relatively soon,
the scale parameter $\theta$ (equivalent to $m_c$) is largely unstable,
and the occurrence of the biggest events makes its value increase.

\begin{figure}[t!]
\begin{center}
\centerline{\includegraphics[trim = 0mm 70mm 0mm
70mm,clip,width=9cm]{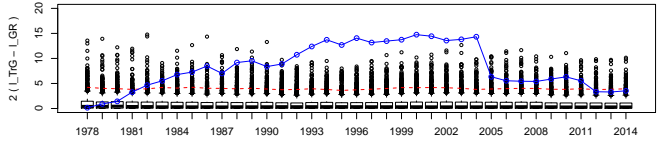}
} \caption{ As Fig. \ref{l_boxplots}, but simulating a power law
with parameter $\beta=0.685$ (Table \ref{taula}) instead of a TrG
distribution. The reshuffling is also as in Fig. \ref{l_boxplots},
as explained in the text.
The 
simulations cannot explain the large empirical values of the log-likelihood-ratio.
\label{figlanueva}}
\end{center}
\end{figure}

{As a complementary control
we invert the situation, simulating
1000 syntetic power-law datasets
with $\beta=0.685$ (Table \ref{taula}),
$a= 5.3 \cdot 10^{17}$ N$\cdot$m,
and $N=6150$, for which the same time reshuffling is perfomed,
in such a way that the order of the order statistics is the same.
In this case, the results of the simulations lead to much smaller values
of the log-ratio, for which the power-law distribution cannot be
rejected (as expected), in contrast with the empirical data,
see Fig. \ref{figlanueva}.}

In summary, the truncated gamma distribution represents the best
alternative to model global shallow earthquake seismic moments, in
comparison with the tapered GR distribution and the power law. The
preponderance of the gamma model is maintained after the occurrence
of the mega-earthquakes taking place from 2004
{and it is only after the 2011 Tohoku earthquake
that it is difficult to decide between power law and TrG.
We have verified that these results are qualitatively similar if
we restrict our study to subduction zones, as defined by
the Flinn-Engdahl's regionalization \cite{Kagan_pageoph99},  
with the main difference that
the values of $l_{trg}-l_{pl}$ become somewhat smaller
and therefore the power-law hypothesis cannot be rejected
after the Tohoku earthquake.}

{In order to reproduce the
time evolution of the statistical results
it suffices that independent gamma seismic moments,
with fixed parameters, are reshuffled
so that the peculiar empirical time correlations of magnitudes are maintained.}
So, although the scale parameter
$\theta$ is not stabilized, and the occurrence or not of more
mega-earthquakes could significantly change its value
\cite{Zoller_grl},
{the current value is enough to explain the
available data}.
It would be very interesting to investigate if
the high values of the likelihood ratio attained before the 2004
Sumatra event could be employed to detect the end of periods of low
global seismic activity. Certainly, more data would be necessary {for that purpose}
\cite{Zoller_grl}.

As an extra argument in favor of the truncated gamma
distribution in front of the tapered GR,
we can bring not a statistical evidence but
physical plausibility; indeed,
the former distribution can be justified as coming from a
branching process that is slightly below its critical point
\cite{Christensen_Moloney,Corral_FontClos}.
{Further reasons that may support
the truncated gamma
are that
this arises
(i) as the maximum entropy outcome
under the constrains of fixed (arithmetic) mean and fixed geometric mean of the seismic moment
\cite{Main_information};
(ii) as the closest to the power law, in terms of the Kullback-Leibler ``distance'',
when the mean seismic moment is fixed \cite{Sornette_Sornette_bssa}; and
(iii) as a stable distribution under a fragmentation process with a power-law
transition rate \cite{Sornette_Sornette_bssa}.}


%


We are grateful to J. del Castillo, Y. Y. Kagan, I. G. Main, and M. Naylor for their feedback.
Research expenses were founded by projects
FIS2012-31324 and MTM2012-31118 from Spanish MINECO,
2014SGR-1307 from AGAUR, and the Collaborative Mathematics
Project from La Caixa Foundation.


%
%

\begin{thebibliography}{10}

\bibitem{Utsu_GR}
T.~Utsu.
\newblock Representation and analysis of earthquake size distribution: a
  historical review and some new approaches.
\newblock {\em Pure Appl. Geophys.}, 155:509--535, 1999.

\bibitem{Malamud_hazards}
B.~D. Malamud.
\newblock Tails of natural hazards.
\newblock {\em Phys. World}, 17 (8):31--35, 2004.

\bibitem{Bak_book}
P.~Bak.
\newblock {\em How Nature Works: The Science of Self-Organized Criticality}.
\newblock Copernicus, New York, 1996.

\bibitem{Knopoff_Kagan77}
L.~Knopoff and Y.~Kagan.
\newblock Analysis of the theory of extremes as applied to earthquake problems.
\newblock {\em J. Geophys. Res.}, 82:5647--5657, 1977.

\bibitem{Corral_Lacidogna}
A.~Corral.
\newblock Scaling and universality in the dynamics of seismic occurrence and
  beyond.
\newblock In A.~Carpinteri and G.~Lacidogna, editors, {\em Acoustic Emission
  and Critical Phenomena}, pages 225--244. Taylor and Francis, London, 2008.

\bibitem{Kagan_pageoph99}
Y.~Y. Kagan.
\newblock Universality of the seismic moment-frequency relation.
\newblock {\em Pure Appl. Geophys.}, 155:537--573, 1999.

\bibitem{Kagan_gji02}
Y.~Y. Kagan.
\newblock Seismic moment distribution revisited: {I}. statistical results.
\newblock {\em Geophys. J. Int.}, 148:520--541, 2002.

\bibitem{Corral_Deluca}
A.~Deluca and A.~Corral.
\newblock Fitting and goodness-of-fit test of non-truncated and truncated
  power-law distributions.
\newblock {\em Acta Geophys.}, 61:1351--1394, 2013.

\bibitem{Kagan_book}
Y.~Y. Kagan.
\newblock {\em Earthquakes: Models, Statistics, Testable Forecasts}.
\newblock Wiley, 2014.

\bibitem{Vere_Jones76}
D.~Vere-Jones.
\newblock A branching model for crack propagation.
\newblock {\em Pure Appl. Geophys.}, 114:711--725, 1976.

\bibitem{Main}
I.~Main.
\newblock Statistical physics, seismogenesis, and seismic hazard.
\newblock {\em Rev. Geophys.}, 34:433--462, 1996.

\bibitem{Zoller_grl}
G.~Z\"oller.
\newblock Convergence of the frequency-magnitude distribution of global
  earthquakes: Maybe in 200 years.
\newblock {\em Geophys. Res. Lett.}, 40:3873--3877, 2013.

\bibitem{Vere_Jones_gji}
D.~Vere-Jones, R.~Robinson, and W.~Yang.
\newblock Remarks on the accelerated moment release model: problems of model
  formulation, simulation and estimation.
\newblock {\em Geophys. J. Int.}, 144(3):517--531, 2001.

\bibitem{Kanamori_77}
H.~Kanamori.
\newblock The energy release in great earthquakes.
\newblock {\em J. Geophys. Res.}, 82(20):2981--2987, 1977.

\bibitem{Kanamori_rpp}
H.~Kanamori and E.~E. Brodsky.
\newblock The physics of earthquakes.
\newblock {\em Rep. Prog. Phys.}, 67:1429--1496, 2004.

\bibitem{Ekstrom2012}
G.~Ekstrom, M.~Nettles, and A.M. Dziewonski.
\newblock The global {CMT} project 2004-2010: Centroid-moment tensors for
  13,017 earthquakes.
\newblock {\em Phys. Earth Planet. Int.}, 200-201:1--9, 2012.

\bibitem{Godano_pingue}
C.~Godano and F.~Pingue.
\newblock Is the seismic moment-frequency relation universal?
\newblock {\em Geophys. J. Int.}, 142:193--198, 2000.

\bibitem{Kagan_tectono10}
Y.~Y. Kagan.
\newblock Earthquake size distribution: Power-law with exponent $\beta\equiv
  1/2$?
\newblock {\em Tectonophys.}, 490:103--114, 2010.

\bibitem{Lay}
T.~Lay.
\newblock Why giant earthquakes keep catching us out.
\newblock {\em Nature}, 483:149--150, 2012.

\bibitem{Main_ng}
I.~G. Main, L.~Li, J.~McCloskey, and M.~Naylor.
\newblock Effect of the {Sumatran} mega-earthquake on the global magnitude
  cut-off and event rate.
\newblock {\em Nature Geosci.}, 1:142, 2008.

\bibitem{Bell_Naylor_Main}
A.~F. Bell, M.~Naylor, and I.~G. Main.
\newblock Convergence of the frequency-size distribution of global earthquakes.
\newblock {\em Geophys. Res. Lett.}, 40:2585--2589, 2013.

\bibitem{Geist_Parsons_NH}
E.~L. Geist and T.~Parsons.
\newblock Undersampling power-law size distributions: effect on the assessment
  of extreme natural hazards.
\newblock {\em Nat. Hazards}, 72:565--595, 2014.

\bibitem{Pawitan2001}
Y.~Pawitan.
\newblock {\em In All Likelihood: Statistical Modelling and Inference Using
  Likelihood}.
\newblock Oxford UP, Oxford, 2001.

\bibitem{Casella}
G.~Casella and R.~L. Berger.
\newblock {\em Statistical Inference}.
\newblock Duxbury, Pacific Grove CA, 2nd edition, 2002.

\bibitem{Gamma_expansion}
{NIST Digital Library of Mathematical Functions}.
\newblock 2014.
\newblock {\tt http://dlmf.nist.gov/8.7\#E3}.

\bibitem{mccullagh1986}
P.~McCullagh and D.~R. Cox.
\newblock Invariants and likelihood ratio statistics.
\newblock {\em Ann. Statist.}, 14(4):1419--1430, 1986.

\bibitem{Self1987}
S.~G. Self and K.-Y. Liang.
\newblock Asymptotic properties of maximum likelihood estimators and likelihood
  ratio tests under nonstandard conditions.
\newblock {\em J. Am. Stat. Assoc.}, 82:605--610, 1987.

\bibitem{Geyer1994}
C.~J. Geyer.
\newblock On the asymptotics of constrained {$M$}-estimation.
\newblock 22(4):1993--2010, 1994.

\bibitem{Kagan_calcutta}
Y.~Y. Kagan.
\newblock Why does theoretical physics fail to explain and predict earthquake
  occurrence?
\newblock In P.~Bhattacharyya and B.~K. Chakrabarti, editors, {\em Modelling
  Critical and Catastrophic Phenomena in Geoscience}, Lecture Notes in Physics,
  705, pages 303--359. Springer, Berlin, 2006.

\bibitem{Kagan_Schoenberg}
Y.~Y. Kagan and F.~Schoenberg.
\newblock Estimation of the upper cutoff parameter for the tapered {Pareto}
  distribution.
\newblock {\em J. Appl. Probab.}, 38A:158--175, 2001.

\bibitem{Vuong}
Q.~H. Vuong.
\newblock Likelihood ratio tests for model selection and non-nested hypotheses.
\newblock {\em Econometrica}, 57(2):307--33, 1989.

\bibitem{Clauset}
A.~Clauset, C.~R. Shalizi, and M.~E.~J. Newman.
\newblock Power-law distributions in empirical data.
\newblock {\em SIAM Rev.}, 51:661--703, 2009.

\bibitem{Christensen_Moloney}
K.~Christensen and N.~R. Moloney.
\newblock {\em Complexity and Criticality}.
\newblock Imperial College Press, London, 2005.

\bibitem{Corral_FontClos}
A.~Corral and F.~Font-Clos.
\newblock Criticality and self-organization in branching processes: application
  to natural hazards.
\newblock In M.~Aschwanden, editor, {\em Self-Organized Criticality Systems},
  pages 183--228. Open Academic Press, Berlin, 2013.

\bibitem{Main_information}
I.~G. Main and P.~W. Burton.
\newblock Information theory and the earthquake frequency-magnitude
  distribution.
\newblock {\em Bull. Seismol. Soc. Am.}, 74(4):1409--1426, 1984.

\bibitem{Sornette_Sornette_bssa}
D.~Sornette and A.~Sornette.
\newblock General theory of the modified {Gutenberg-Richter} law for large
  seismic moments.
\newblock {\em Bull. Seismol. Soc. Am.}, 89(4):1121--1130, 1999.

\end{thebibliography}
%


\end{document}